\documentclass[aps,nofootinbib]{revtex4-2}
\usepackage{amssymb}
\usepackage{eurosym}
\usepackage{amsfonts}
\usepackage{geometry}
\usepackage{color}
\usepackage{ulem}
\usepackage{bbm}
\usepackage{color}

\usepackage{mathtools}
\begin{document}

\title{Generalized Extended Uncertainty Principles, Liouville theorem and
density of states: Snyder-de Sitter and Yang models}

\author{A. Pacho\l}

\affiliation{Department of Microsystems, University of South-Eastern Norway, Campus Vestfold, Norway}

\begin{abstract}
Modifications in quantum mechanical phase space lead to the changes in the Heisenberg uncertainty principle, which can result in the Generalized
Uncertainty Principle (GUP) or the Extended Uncertainty Principle (EUP), introducing quantum gravitational effects at small and large distances, respectively. A combination of GUP and EUP, the Generalized Extended 
Uncertainty Principle (GEUP or EGUP), further generalizes these modifications by incorporating noncommutativity in both coordinates and
momenta. This paper examines the impact of GEUP 
on the analogue of the Liouville theorem in statistical physics and density of states within the classical limit of
non-relativistic quantum mechanics framework. We find a weighted phase space volume element, invariant under the infinitesimal time evolution, in the cases of
Snyder-de Sitter and Yang models, presenting how GEUP alters the density of
states, potentially affecting physical (thermodynamical) properties. Special cases, obtained in certain limits from the above models are discussed. New higher order types of GEUP and EUP are also proposed. 

\end{abstract}

\maketitle

\section{Introduction}

The search for the fundamental theory of Quantum Gravity has been supported
by the development of phenomenological models that explore possible
modifications to the known quantum mechanical or gravitational phenomena. In
this context the various purely phenomenological proposals have appeared
aiming to capture potential signatures of quantum effects of gravity, hinting towards
possible experimental set-ups which would help guide the way in the full
formulation of the theory. What is known is that one needs to challenge the
concepts of classical space-time. The idea that the structure of space-time
should be modified is a common feature of various approaches to Quantum
Gravity, such as String Theory, Loop Quantum Gravity, Causal Dynamical
Triangulations, Asymptotically Safe Quantum Gravity, Horava-Lifshitz Gravity
and Noncommutative Geometry, just to name a few \cite
{Seiberg:1999vs,Gross:1987ar,Rovelli:1994ge,Ambjorn:2005db,Lauscher:2005qz,Horava:2009if}. Many of these share also the concept of minimum length \cite{Garay:1994en}. Such minimal length can be implemented in quantum mechanics through
introducing the modifications in Uncertainty Principles (UP)\footnote{For simplicity, to illustrate the main points, we use 1-dimensional case
throughout the introduction.}: 
\begin{equation}
\Delta x\Delta p\geq \frac{1}{2}\left\vert \langle \left[ x,p\right] \rangle
\right\vert   \label{UP}
\end{equation}%
where $\Delta $ denotes standard deviation and $\langle \quad \rangle $ the quantum
expectation value on a
given state.
From the above relation, it is clear that any modifications in the canonical Heisenberg commutation relations between coordinates and momenta will imply changes in the UP. Various generalizations of UP have been considered in the literature. The most common type of such modifications is the Generalized Uncertainty Principle (GUP) with the momenta-dependent right hand side of \eqref{UP}. It was firstly proposed in \cite{Maggiore:1993rv} and then related to the specific algebraic structure of quantum phase space \cite{Maggiore:1993zu,Maggiore:1993kv,Kempf:1994su,Chang:2001bm}. The right hand side of \eqref{UP} may include quadratic terms in momenta (resulting in QGUP), see e.g. \cite{Kempf:1994su,Chang:2001bm,Chang:2001kn} or have linear and quadratic terms
(LQGUP), see e.g. \cite{Vagenas:2019wzd,Bernaldez:2022muh, Wojnar:2023bvv, Wojnar:2024hcg, Ali:2024tbd}. In momenta-dependent GUP relations one
introduces the parameter, usually denoted by $\beta $, which is related to
the Planck (length) scale $l_{P}$, hence linking GUPs with Planck scale
physics and quantum gravitational effects. Such phenomenological models have
attracted a lot of attention 
\cite
{Maggiore:1993zu,Maggiore:1993kv,Kempf:1994su,Chang:2001bm,
Chang:2001kn,Vagenas:2019wzd,Bernaldez:2022muh,Wojnar:2023bvv, Wojnar:2024hcg,Ali:2024tbd,Scardigli:1999jh,Capozziello:1999wx,Brau:2006ca,Das:2008kaa,Scardigli:2016pjs,Harikumar:2017suv,Wang:2010ct,Ali:2011ap,Ali:2013ii,Bawaj:2014cda,Mathew:2017drw,Tamburini:2021inp,Das:2021lrb,Tunacao:2022ffq,Campbell:2023dwm,Pachol:2023tqa,Pachol:2023bkv}, see also \cite{Bosso:2023aht} for a recent review and more references on
the topic.

On the other hand, the symmetry between the position and momenta in the canonical (quantum mechanical) commutation relations, as well as Born reciprocity \cite{Born:1949yva}, suggests the possibility of introducing corrections to the
Heisenberg uncertainty principle by including the modifications proportional
to coordinates instead, i.e. so that the RHS of \eqref{UP} is quadratic or
linear in coordinates instead of momenta. This complimentary type of uncertainty relation has been called an Extended
Uncertainty Principle (EUP) \cite{Bolen:2004sq,Park:2007az}. Here,
the parameter of the model, usually denoted by $\alpha $, is related with the
non-vanishing cosmological constant $\Lambda $ and in this way it can be
embedded in the non-relativistic quantum mechanics. 
It has been shown \cite{Mignemi:2009ji} that this type of modification to
the UP is related with the (Anti-)de Sitter geometric background, and the
parameter $\alpha $ is then naturally linked with the (Anti-)de Sitter
radius. While GUP, with $\beta \sim l_{P}$, exposes the gravitational
modifications in quantum mechanics at the small distances; EUP, with $\alpha
\sim \Lambda $ introduces the idea of modifications at large distances. 
Relying on the analogy with GUPs, various types of such coordinate-dependent models
have been introduced leading to different interesting physical effects, see e.g. \cite{Bolen:2004sq,Park:2007az,Hamil:2019pum,Nozari:2024wir}

Combination of GUP and EUP (i.e. considering both coordinate and momenta
dependent terms on the RHS of \eqref{UP}) appeared firstly in the
construction of noncommutative quantum mechanics with quantum groups as a
symmetry \cite{Kempf:1993bq}, where both nonzero minimal uncertainties have appeared for position and momentum observables, respectively. Later on
the relation of the type: 
\begin{equation}
\Delta x\Delta p\geq \frac{\hbar }{2}(1+\alpha ^{2}(\Delta x)^{2}+\beta
^{2}(\Delta p)^{2})  \label{egup}
\end{equation}
under the name of Generalized Extended Uncertainty Principle
(GEUP, or alternatively EGUP), has been studied in various contexts, see e.g. \cite{Bolen:2004sq,Park:2007az,Mignemi:2009ji,Bambi:2007ty}, for more recent work
see e.g. \cite{Wagner:2022rjg} where Born reciprocity is also promoted or see e.g. \cite{Moulla:2024qft} for applications in statistical physics.

Even though the origins of GUPs, EUPs and GEUPs can be tracked back to the
noncommutative geometry and quantum groups as the underlying mathematical
frameworks \cite{Maggiore:1993kv,Kempf:1993bq}, these models are mainly
treated as purely phenomenological in the majority of the literature related
to this subject. They have been used (and have been quite fruitful) in
providing predictions for various phenomenological effects. However, such departure
from the underlying mathematical framework resulting in the specific form of
GUP, EUP or GEUPs together with the non-uniqueness of the defining
commutation relations between coordinates and momenta has lead to many
conceptual shortcomings and ambiguities (see e.g. the recent review \cite{Bosso:2023aht} or \cite{Raghavi:2024fpr} for discussion of some of the arising issues). Moreover, it is worth to point out that the minimal length may
not necessarily appear in such models if one bases only on the modifications of the quantum mechanical phase space, see e.g. \cite
{Segreto:2022clx,Bishop:2019yft,Bishop:2022vzr,Bosso:2023sxr}.

In this paper, following the point of view that the noncommutative geometry
is the underlying framework (mathematical language) of the possible
fundamental theory, we assume that the quantization process of general
relativity, includes the quantization of space-time \cite%
{Doplicher:1994tu,Doplicher:1994zv}, i.e. requiring the space-time
coordinates to become noncommutative $[\hat{x}_{\mu },\hat{x}_{\nu }]\neq 0$. The noncommutativity of space-time introduces corrections to the canonical
quantum-mechanical phase space relations. Hence modifications to the UP
appear as a natural consequence. In this view, quadratic GUPs have been
mainly linked with the Snyder model \cite%
{Maggiore:1993kv,Pachol:2023tqa,Pachol:2023bkv}, with noncommuting
coordinates and commuting momenta. EUPs, on the other hand can be linked
with the (anti)-de Sitter geometric background and algebraic structure with
commuting coordinates but noncommuting momenta \cite{Mignemi:2009ji}. The
main aim of this paper is to show how the noncommutativity of both coordinates $[\hat{x}_i,\hat{x}_j]\neq 0$
and momenta $[\hat{p}_i,\hat{p}_j]\neq 0$ leading to GEUPs (and in some cases, the appearance of the
nonzero minimal uncertainties in positions and momenta separately)
affects the density of states and the analogue of the Liouville theorem in
statistical mechanics. For this reason we limit ourselves to the case of
non-relativistic quantum mechanics. We find the new form of weighted
phase space volume element in the presence of GEUPs in the specific cases of
Snyder-de Sitter (SdS) and Yang models. We show that noncommuting coordinates and noncommuting momenta with the corresponding GEUP require
introducing the modification in the density states and this may impact
various physical effects and thermodynamical properties (as it has been shown in the case of various GUP models).

In the next section, we summarise the framework generalized to the case
where both coordinates and momenta do not commute and present the general
formulae for the Jacobian arising from the variable change under the
infinitesimal time evolution. In Sec. 3, we specify the model to the Snyder-de
Sitter (SdS) algebra, as only then 
we can identify the required weight factor (which depends on both
coordinates and momenta) for the phase space volume element. 
The factor we obtain in this case does not depend on the space dimension $D$. We also
discuss the special cases obtained in the certain limits of the parameters,
giving GUP or EUP relations (by reducing the starting SdS algebra to Snyder
or dS algebras, respectively) and we give the expressions for the
phase space volume element in these cases. In Sec. 4, we discuss the Yang model
and consider specific realizations of its generators on the canonical phase
space leading to another type for (higher order) GEUP relation. The weighted phase space
volume element is obtained for this case in 1 dimension by adapting the
result from SdS model. Special cases, obtained in certain limits are also discussed, one provides the known "square-root" (or Maggiore) GUP and the other leads to the new higher order ("square-root") type of EUP. In Sec. 5, the Lie algebraic case with commuting momenta is briefly considered with the fuzzy sphere as an example. It is shown that in this case the phase space volume element stays invariant under the time evolution and there is no change in the density of states.

\section{Preliminaries}

In this section we set up the most general framework for investigating the
effects of noncommuting coordinates and noncommuting momenta (and in
principle the appearance of the nonzero length and momentum uncertainties)
on the density of states in the phase space so that we can adapt the
Liouville theorem in statistical physics to this new scenario. For this reason we do not fix the specific choice
for the noncommutativity of space-time or momena and the resulting
deformation of the quantum phase space commutation relations at this point
yet. 

We start with the following (most general) set of commutation relations
describing the noncommutative quantum mechanical phase space algebra: 
\begin{equation}
\lbrack \hat{x}_{i},\hat{x}_{j}]=i\hbar a_{ij}\left( \hat{x},\hat{p}\right)
,\qquad \lbrack \hat{p}_{i},\hat{p}_{j}]=i\hbar b_{ij}\left( \hat{x},\hat{p}%
\right)   \label{genCRhat}
\end{equation}%
\begin{equation}
\lbrack \hat{x}_{i},\hat{p}_{j}]=i\hbar c_{ij}\left( \hat{x},\hat{p}\right) ,
\label{genCRxp}
\end{equation}%
where $a_{ij}\left( \hat{x},\hat{p}\right) ,b_{ij}\left( \hat{x},\hat{p}%
\right) ,c_{ij}\left( \hat{x},\hat{p}\right) $ are functions which may
include all kinds of terms (linear, quadratic, higher order etc.) in
space-time coordinates $\hat{x}$ and momenta $\hat{p}$, such that the Jacobi
identities are satisfied. We will focus on the non-relativistic general case
in any dimension $D$ with $i,j=1,2,\ldots ,D$. When none of the above commutators are zero (see e.g. \cite{Kowalski-Glikman:2004fso,Yang:1947ud,Kowalski-Glikman:2003qjp,Kowalski-Glikman:2004fso,Wagner:2022rjg,Lukierski:2023gxf}), this will lead to the interesting types of the GEUPs: $\Delta \hat{x}_i\Delta \hat{p}_j\geq \frac{\hbar}{2}\left\vert \langle c_{ij}(\hat{x},\hat{p}) \rangle
\right\vert$ with the specific form of RHS depending on the concrete choice of the algebra \eqref{genCRhat}, \eqref{genCRxp}. 

In the phenomenological approaches, one considers the right hand side of %
\eqref{genCRxp} as the definition of a new effective value of $\hbar $ which
(in the most general case) may depend on both coordinates and momenta. This
means that the size of the unit cell that each quantum state occupies in the
phase space can be thought of as being also coordinates and momenta
dependent. This will have an effect on the density of states and as a
consequence affect physical, for example thermodynamical, properties. 
For this interpretation to be valid, the volume of phase space must evolve
in such a way that the number of states does not change with time, in other
words we are are looking for the analogue of the Liouville theorem in
statistical physics. Similar investigation has been done in the case of the
quadratic GUP \cite{Chang:2001bm,Chang:2001kn} or GUPs with higher order terms \cite{Pedram:2012my}. Note that here we assume that both coordinates and momenta do
not commute. Nevertheless, later on we shall see how this more general case
can be reduced to the special cases like the GUP or EUP (which include some
commutative generators).

Starting with the quantum mechanical commutation relations \eqref{genCRhat}, %
\eqref{genCRxp}, these will correspond to the Poisson brackets in classical
mechanics: 
\begin{equation}
\frac{1}{i\hbar }[\hat{A},\hat{B}]\xrightarrow{\tiny{classical\ limit}}%
\{A,B\}.
\end{equation}%
Therefore, in the classical limit\footnote{%
It is worth to note that the classical limit in GUP models may be more
involved than just applying the above transformation, see e.g. \cite%
{Casadio:2020rsj} where a possible way out is suggested to derive the GUP
relations from the (explicitly state dependent) deformed commutators between
coordinates and momenta. Here, on the contrary to the phenomenological GUP
approaches, we are starting from the noncommutative model as a possible
description of the quantization of space-time, hence the starting algebra is
fixed from the beginning and modified uncertainty relations and all effects
are a consequence of the starting choice of quantum-deformed phase space.}
we obtain (in our shortcut notation): 
\begin{equation}
\{x_{i},x_{j}\}=a_{ij}(x,p),\qquad \{p_{i},p_{j}\}=b_{ij}(x,p),  \label{xx}
\end{equation}%
\begin{equation}
\{x_{i},p_{j}\}=c_{ij}(x,p).  \label{xp}
\end{equation}%
The time evolution of the coordinates and momenta is governed by the equations (where the summation convention is assumed) 
\begin{eqnarray}
\dot{x}_{i} &=&\{x_{i},H\}=\{x_{i},p_{j}\}\frac{\partial H}{\partial p_{j}}%
+\{x_{i},x_{j}\}\frac{\partial H}{\partial x_{j}}=c_{ij}\frac{\partial H}{%
\partial p_{j}}+a_{ij}\frac{\partial H}{\partial x_{j}}, \\
\dot{p}_{i} &=&\{p_{i},H\}=-\{x_{j},p_{i}\}\frac{\partial H}{\partial x_{j}}%
+\{p_{i},p_{j}\}\frac{\partial H}{\partial p_{j}}\delta t=-c_{ji}\frac{%
\partial H}{\partial x_{j}}+b_{ij}\frac{\partial H}{\partial p_{j}}.
\end{eqnarray}%
The Liouville theorem requires that the phase space volume is
invariant under time evolution. Hence we consider an infinitesimal time
interval $\delta t$ and the evolution of the coordinates and momenta during $%
\delta t$ is: 
\begin{equation*}
x_{i}^{\prime }=x_{i}+\delta x_{i},\quad p_{i}^{\prime }=p_{i}+\delta p_{i}
\end{equation*}%
where 
\begin{equation}
\delta x_{i}=\dot{x}_{i}\delta t=\left( c_{ij}\frac{\partial H}{\partial
p_{j}}+a_{ij}\frac{\partial H}{\partial x_{j}}\right) \delta t,\qquad \delta
p_{i}=\dot{p}_{i}\delta t=\left( -c_{ji}\frac{\partial H}{\partial x_{j}}%
+b_{ij}\frac{\partial H}{\partial p_{j}}\right) \delta t.  \label{delp}
\end{equation}
The infinitesimal phase space volume after this infinitesimal time evolution
will be: 
\begin{equation*}
d^{D}x^{\prime }d^{D}p^{\prime }=Jd^{D}xd^{D}p
\end{equation*}%
with the Jacobian 
\begin{equation}
J=\left\vert \frac{\partial \left( x_{1}^{\prime },\ldots ,x_{D}^{\prime
},p_{1}^{\prime },\ldots ,p_{D}^{\prime }\right) }{\partial \left(
x_{1},\ldots ,x_{D},p_{1},\ldots ,p_{D}\right) }\right\vert =1+\left( \frac{%
\partial \delta x_{i}}{\partial x_{i}}+\frac{\partial \delta p_{i}}{\partial
p_{i}}\right) +\ldots 
\end{equation}%
where we used: 
\begin{eqnarray}
\frac{\partial x_{i}^{\prime }}{\partial x_{j}} &=&\delta _{ij}+\frac{%
\partial \delta x_{i}}{\partial x_{j}},\quad \frac{\partial x_{i}^{\prime }}{%
\partial p_{j}}=\frac{\partial \delta x_{i}}{\partial p_{j}}, \\
\frac{\partial p_{i}^{\prime }}{\partial x_{j}} &=&\frac{\partial \delta
p_{i}}{\partial x_{j}},\quad \frac{\partial p_{i}^{\prime }}{\partial p_{j}}%
=\delta _{ij}+\frac{\partial \delta p_{i}}{\partial p_{j}}.
\end{eqnarray}%
Up to the first order in $\delta t$ (based on \eqref{delp})
we get: 
\begin{eqnarray}
\left( \frac{\partial \delta x_{i}}{\partial x_{i}}+\frac{\partial \delta
p_{i}}{\partial p_{i}}\right)  &=&\frac{\partial }{\partial x_{i}}\left(
c_{ij}\frac{\partial H}{\partial p_{j}}+a_{ij}\frac{\partial H}{\partial
x_{j}}\right) \delta t+\frac{\partial }{\partial p_{i}}\left( -c_{ji}\frac{%
\partial H}{\partial x_{j}}+b_{ij}\frac{\partial H}{\partial p_{j}}\right)
\delta t  \notag \\
&=&\left[ \left( \frac{\partial }{\partial x_{i}}a_{ij}-\frac{\partial }{%
\partial p_{i}}c_{ji}\right) \frac{\partial H}{\partial x_{j}}+\left( \frac{%
\partial }{\partial x_{i}}c_{ij}+\frac{\partial }{\partial p_{i}}%
b_{ij}\right) \frac{\partial H}{\partial p_{j}}\right] \delta t
\end{eqnarray}%
where the the terms with mixed derivatives have cancelled each other and the
antisymetricity of the Poisson brackets was used to cancel the remaining
terms. 
This way we find the expression for the time evolved infinitesimal phase
space volume (in the first order of $\delta t$) as: 
\begin{equation}
d^{D}x^{\prime }d^{D}p^{\prime }=d^{D}xd^{D}p\left( 1+\left[ \left( \frac{%
\partial }{\partial x_{i}}a_{ij}-\frac{\partial }{\partial p_{i}}%
c_{ji}\right) \frac{\partial H}{\partial x_{j}}+\left( \frac{\partial }{%
\partial x_{i}}c_{ij}+\frac{\partial }{\partial p_{i}}b_{ij}\right) \frac{%
\partial H}{\partial p_{j}}\right] \delta t+O\left( \delta t^{2}\right)
\right)   \label{dxdp}
\end{equation}%
valid for any Poisson algebra \eqref{xx}, \eqref{xp} obtained as the
classical limit of any noncommutative model \eqref{genCRhat}, \eqref{genCRxp}%
. Since in general the terms in the brackets will not cancel out, already
here we see the need to introduce the weight to the phase space volume
element so that the analogue of the Liouville theorem is satisfied. The
specific factor has to be chosen in such a way that the weighted phase space
volume is invariant under the time evolution, i.e. so that:
\begin{equation}
\frac{d^{D}x^{\prime }d^{D}p^{\prime }}{F\left( x^{\prime },p^{\prime
}\right) }\sim \frac{d^{D}xd^{D}p}{F\left( x,p\right) }.  \label{dxdpF}
\end{equation}%
It is straightforward to notice from \eqref{dxdp} that the noncanonical Poisson brackets for
coordinates and for momenta \eqref{xx}, as well as the mixed relation between
coordinates and momenta \eqref{xp} are crucial in the choice of
the weighted phase space volume. Hence the underlying noncommutative
geometry (and the choice of \eqref{genCRhat}, \eqref{genCRxp}) is the
intrinsic feature of phenomenological models investigating the possible
effects arising from such modified phase space volume element (and
consequently affecting the thermodynamical properties of physical systems).
To be able to investigate the time evolution of the weight factor $F(x,p)$, we need to consider the concrete
noncommutative model, which we shall do in the next section.

\section{Snyder-de Sitter model and the density of states}

Snyder-de Sitter (SdS) model, which includes the noncommutative space-time
coordinates and the noncommutative momenta, was proposed \cite%
{Kowalski-Glikman:2004fso} as a generalization of the Snyder model \cite%
{snyder1947quantized} to a space-time background of constant curvature and
it was investigated in many contexts, see e.g. 
\cite{Mignemi:2011wh}, 
\cite{Banerjee:2011ag}. 
In SdS model, the noncommutativity among space-time coordinates corresponds to the curved
momentum space, and vice-versa noncommutative momenta lead to the curved
space-time. Since we are interested in investigating how the
noncommutativity of both coordinates and momenta affects the density of
states we choose SdS set of commutation relations which exhibit these features:
\begin{equation}
\lbrack \hat{x}_{\mu },\hat{x}_{\nu }]=i\hbar \beta ^{2}\hat{M}_{\mu \nu
},\qquad \lbrack \hat{p}_{\mu },\hat{p}_{\nu }]=i\hbar \alpha ^{2}\hat{M}%
_{\mu \nu },  \label{xxTSR}
\end{equation}%
\begin{equation}
\lbrack \hat{M}_{\mu \nu },\hat{x}_{\rho }]=i\hbar (\eta _{\mu \rho }\hat{x}%
_{\nu }-\eta _{\nu \rho }\hat{x}_{\mu }),  \label{Mx}
\end{equation}%
\begin{equation}
\lbrack \hat{M}_{\mu \nu },\hat{p}_{\rho }]=i\hbar (\eta _{\mu \rho }\hat{p}%
_{\nu }-\eta _{\nu \rho }\hat{p}_{\mu }),  \label{Mp}
\end{equation}%
\begin{equation}
\lbrack \hat{M}_{\mu \nu },\hat{M}_{\rho \tau }]=i\hbar (\eta _{\mu \rho }%
\hat{M}_{\nu \tau }-\eta _{\mu \tau }\hat{M}_{\nu \rho }+\eta _{\nu \tau }%
\hat{M}_{\mu \rho }-\eta _{\nu \rho }\hat{M}_{\mu \tau }),  \label{MM}
\end{equation}%
with the modified quantum mechanical phase space relation: 
\begin{equation}
\lbrack \hat{x}_{\mu },\hat{p}_{\nu }]=i\hbar \left( \eta _{\mu \nu }+\alpha
^{2}\hat{x}_{\mu }\hat{x}_{\nu }+\beta ^{2}\hat{p}_{\mu }\hat{p}_{\nu
}+\alpha \beta (\hat{x}_{\mu }\hat{p}_{\nu }+\hat{p}_{\mu }\hat{x}_{\nu }-%
\hat{M}_{\mu \nu })\right).  \label{xpTSR}
\end{equation}%
%
%
Here $\eta _{\mu \nu }$ is the flat metric with Lorentzian signature and $%
\mu ,\nu =0,1,\ldots ,D$. Depending on the sign\footnote{%
For the Jacobi identities to be satisfied both coupling constants $\alpha
^{2}$ and $\beta ^{2}$ must have the same sign. Following the convention of 
\cite{Mignemi:2011wh} when $\alpha ^{2}$ and $\beta ^{2}$ are negative then $%
\alpha \beta $ is negative.} of $\alpha ^{2}$ the Lorentz generators $\hat{M}%
_{\mu \nu }$ and momenta $\hat{p}_{\mu }$ generate de Sitter (dS) 
or Anti-de Sitter (AdS) 
subalgebras. 
This model involves, besides the speed of light\footnote{%
We set $c=1$ throughout the paper.}, two other observer-independent constants%
\footnote{%
SdS is also known as doubly special relativity (DSR) in de Sitter space \cite%
{Kowalski-Glikman:2003qjp} or triply special relativity (TSR) model \cite%
{Kowalski-Glikman:2004fso}.}, the Planck length 
as well as the de Sitter radius which is related to the
cosmological constant
. More
precisely, the parameter $|\alpha ^{2}|$ has the dimension of
the inverse of the square of length and it can be identified with the
(Anti-)de Sitter radius and the cosmological constant as its inverse,
while the parameter $|\beta ^{2}|$ has the dimension of the inverse square
of mass and it can be is identified with $1/M_{P}^{2}=l_{P}^{2}$, where $l_P$ and $M_{P}$ are the Planck length and mass, respectively. In general, $\alpha ^{2}$ and $\beta ^{2}$ can
take positive or negative value, which generates models with very different
properties \cite{Mignemi:2011wh}.

In the following we focus on the non-relativistic quantum mechanical
counterpart (which in $D=3$ would give the Snyder model restricted to a
three-dimensional sphere) however we keep unspecified dimension $D$, $%
i,j=1,2,\ldots ,D$. The GEUP corresponding to \eqref{xpTSR} has the form
similar to \eqref{egup} \cite{Mignemi:2009ji} \footnote{%
Note that the same GEUP can be realized by a different deformed Heisenberg
algebras, see e.g. \cite{Wagner:2022rjg}.}, but additional (mixed) terms may
appear on the RHS depending on the choice of the representation of the
angular momentum used, see e.g. \cite{Mignemi:2011wh}. Without any
additional assumptions we can, at most, write:
\begin{equation}
\left( \Delta \hat{x}_{i}\right) \left( \Delta \hat{p}_{i}\right) \geq \frac{\hbar }{2}%
\left\vert 1+\alpha ^{2}\left( \Delta \hat{x}\right) ^{2}+\beta ^{2}\left( \Delta
\hat{p}\right) ^{2}+\gamma \right\vert 
\end{equation}
where we used $\ \left( \Delta \hat{x}\right) ^{2}=\left\langle \hat{x}%
^{2}\right\rangle -\left\langle \hat{x}\right\rangle ^{2}$ , $\left( \Delta
\hat{p}\right) ^{2}=\left\langle \hat{p}^{2}\right\rangle -\left\langle \hat{p}%
\right\rangle ^{2}$ and $\gamma =\alpha ^{2}\left\langle \hat{x}%
\right\rangle ^{2}+\beta ^{2}\left\langle \hat{p}\right\rangle ^{2}+\alpha
\beta (\left\langle \hat{x}\hat{p}\right\rangle +\left\langle \hat{p}\hat{x}%
\right\rangle )$ as well as the standard notation $\hat{x}^{2}=\hat{x}_{i}%
\hat{x}_{i}\,$, $\hat{p}^{2}=\hat{p}_{i}\hat{p}_{i}$, $\hat{x}\hat{p}=\hat{x}%
_{i}\hat{p}_{i}$ was assumed, while on the LHS 
there is no summation. In 1-dimensional case, one can show \cite{Mignemi:2011wh} that, for both $\alpha ^{2}>0$ and $\beta ^{2}>0$ the
bounds on the localization in position and momentum space arise, while when $
\alpha ^{2}<0$ and $\beta ^{2}<0$ a combination of spatial and momentum
coordinates becomes bounded instead.

The classical (non-relativistic) Snyder-de Sitter Poisson algebra is given by\footnote{where we used the fact that the above algebra (and relation \eqref{xpTSR})
can be implemented on the canonical phase space 
with Lorentz generators defined as $M_{\mu \nu }=x_{\mu }p_{\nu }-x_{\nu
}p_{\mu }$.}: 
\begin{eqnarray}
\{x_{i},x_{j}\} &=&\beta ^{2}(x_{i}p_{j}-x_{j}p_{i})\equiv {a}_{ij},\qquad
\{p_{i},p_{j}\}=\alpha ^{2}(x_{i}p_{j}-x_{j}p_{i})\equiv {b}_{ij},
\label{SdSP} \\
\{x_{i},p_{j}\} &=&\delta _{ij}+\alpha ^{2}x_{i}x_{j}+\beta
^{2}p_{i}p_{j}+2\alpha \beta p_{i}x_{j}\equiv c_{ij}.  \notag
\end{eqnarray}
and supplemented by the relations with Lorentz generators. By using the general
set up from the previous section and now the explicit form of SdS model %
\eqref{SdSP} we obtain that the infinitesimal phase space volume element after the infinitesimal
time evolution as: 
\begin{equation}
d^{D}x^{\prime }d^{D}p^{\prime }=d^{D}xd^{D}p\left( 1+2\left[ \left( \alpha
^{2}x_{j}+\alpha \beta p_{j}\right) \frac{\partial H}{\partial p_{j}}-\left(
\beta ^{2}p_{j}+\alpha \beta x_{j}\right) \frac{\partial H}{\partial x_{j}}%
\right] \delta t+O\left( \delta t^{2}\right) \right)   \label{jacobian}
\end{equation}%
where we explicitly used \eqref{SdSP} in \eqref{dxdp} with the following: 
\begin{eqnarray}
\frac{\partial }{\partial x_{i}}{a}_{ij} &=&\beta ^{2}(D-1)p_{j},\qquad 
\frac{\partial }{\partial x_{i}}c_{ij}=\alpha ^{2}\left( D+1\right)
x_{j}+2\alpha \beta p_{j},  \label{partial1} \\
\frac{\partial }{\partial p_{i}}{b}_{ij} &=&\alpha ^{2}(1-D)x_{j}=-\alpha
^{2}(D-1)x_{j},\qquad \frac{\partial }{\partial p_{i}}c_{ji}=\beta
^{2}\left( D+1\right) p_{j}+2\alpha \beta x_{j}.  \label{partial2}
\end{eqnarray}%
In the remaining part of this section we will show that for the SdS Poisson
algebra \eqref{SdSP} the analogue of the Liouville theorem is satisfied when
we consider the following weighted phase space volume 
\begin{equation}
\frac{d^{D}xd^{D}p}{\left( 1+\alpha ^{2}x^{2}+\beta ^{2}p^{2}+2\alpha \beta
x\cdot p\right) }  \label{dxdp_inv}
\end{equation}%
which is invariant under the infinitesimal time evolution. \newline
To show this, we consider the time evolution of each of the terms in the
proposed factor $F(x,p)=1+\alpha ^{2}x^{2}+\beta ^{2}p^{2}+2\alpha \beta
x\cdot p$ during the infinitesimal time interval $\delta t$. At first we
keep all the formulae as general as possible and express them in terms of $%
a_{ij}$, $b_{ij}$ and $c_{ij}$ and only at the end we will specify to SdS
case \eqref{SdSP}. We get the following expressions: 
\begin{eqnarray}
x^{\prime 2} &=&\left( x_{i}+\delta x_{i}\right) ^{2}  \notag \\
&=&x^{2}+2x_{i}\delta x_{i}+O\left( \delta t^{2}\right)   \notag \\
&=&x^{2}+2x_{i}\left( c_{ij}\frac{\partial H}{\partial p_{j}}+a_{ij}\frac{%
\partial H}{\partial x_{j}}\right) \delta t+O\left( \delta t^{2}\right) ,
\end{eqnarray}
\begin{eqnarray}
p^{\prime 2} &=&\left( p_{i}+\delta p_{i}\right) ^{2}  \notag \\
&=&p^{2}+2p_{i}\delta p_{i}+O\left( \delta t^{2}\right)   \notag \\
&=&p^{2}+2p_{i}\left( -c_{ji}\frac{\partial H}{\partial x_{j}}+b_{ij}\frac{%
\partial H}{\partial p_{j}}\right) \delta t+O\left( \delta t^{2}\right) ,
\end{eqnarray}%
and for the last term 
\begin{eqnarray}
x^{\prime }\cdot p^{\prime } &=&x^{\prime }p^{\prime }=\left( x_{i}+\delta
x_{i}\right) \left( p_{i}+\delta p_{i}\right)   \notag \\
&=&xp+p_{i}\delta x_{i}+x_{i}\delta p_{i}+O\left( \delta t^{2}\right)  
\notag \\
&=&xp+p_{i}\left( c_{ij}\frac{\partial H}{\partial p_{j}}+a_{ij}\frac{%
\partial H}{\partial x_{j}}\right) \delta t+x_{i}\left( -c_{ji}\frac{%
\partial H}{\partial x_{j}}+b_{ij}\frac{\partial H}{\partial p_{j}}\right)
\delta t+O\left( \delta t^{2}\right) .
\end{eqnarray}%
%
%
Therefore the time evolution of the whole factor $F(x,p)$ can be first
written generally as (where for simplicity we use $2\alpha \beta =\gamma $): 
\begin{eqnarray}
1+\alpha ^{2}x^{\prime 2}+\beta ^{2}p^{\prime 2}+\gamma x^{\prime }p^{\prime
} &=&1+\alpha ^{2}x^{2}+\beta ^{2}p^{2}+\gamma xp+  \notag \\
&&+\left[ 2\alpha ^{2}x_{i}c_{ij}+2\beta ^{2}p_{i}b_{ij}+\gamma \left(
p_{i}c_{ij}+x_{i}b_{ij}\right) \right] \frac{\partial H}{\partial p_{j}}%
\delta t+  \notag \\
&&+\left[ 2\alpha ^{2}x_{i}a_{ij}-2\beta ^{2}p_{i}c_{ji}+\gamma \left(
p_{i}a_{ij}-x_{i}c_{ji}\right) \right] \frac{\partial H}{\partial x_{j}}%
\delta t+O\left( \delta t^{2}\right) .
\end{eqnarray}%
Now specialising this to the case of SdS \eqref{SdSP}, after plugging in the
expressions for $a_{ij}$, $b_{ij}$ and $c_{ij}$, we obtain (in the first
order of $\delta t$): {\small
\begin{eqnarray}
&&1+\alpha ^{2}x^{\prime 2}+\beta ^{2}p^{\prime 2}+\gamma x^{\prime
}p^{\prime }=1+\alpha ^{2}x^{2}+\beta ^{2}p^{2}+\gamma xp  \notag \\
&&+\left[ 2\alpha ^{2}\left( x_{j}+\alpha ^{2}x^{2}x_{j}+\beta ^{2}\left(
xp\right) p_{j}+\gamma \left( xp\right) x_{j}\right) +2\beta ^{2}\alpha
^{2}\left( xp\right) p_{j}-2\beta ^{2}\alpha ^{2}x_{j}p^{2}+\gamma \left(
p_{j}+\alpha ^{2}x^{2}p_{j}+\beta ^{2}p^{2}p_{j}+\gamma p^{2}x_{j}\right) %
\right] \frac{\partial H}{\partial p_{j}}\delta t+  \notag \\
&&+\left[ -2\beta ^{2}\left( p_{j}+\alpha ^{2}x_{j}\left( xp\right) +\beta
^{2}p_{j}p^{2}+\gamma \left( xp\right) p_{j}\right) +2\alpha ^{2}\beta
^{2}x^{2}p_{j}-2\alpha ^{2}\beta ^{2}x_{j}\left( xp\right) -\gamma \left(
x_{j}+\alpha ^{2}x^{2}x_{j}+\beta ^{2}p^{2}x_{j}+\gamma x^{2}p_{j}\right) %
\right] \frac{\partial H}{\partial x_{j}}\delta t.  \notag \\
&&
\end{eqnarray}
}
\normalsize 
Since the aim is to factorise the whole expression $(1+\alpha
^{2}x^{2}+\beta ^{2}p^{2}+\gamma xp)$ on the right hand side of this
equality, we need to rearrange all the terms in the square brackets in such
a way so that we can recognize the whole factor $F(x,p)$%
. In this way, we obtain (once we returned to the notation $\gamma =2\alpha
\beta $): 
\begin{eqnarray}
&&1+\alpha ^{2}x^{\prime 2}+\beta ^{2}p^{\prime 2}+2\alpha \beta x^{\prime
}p^{\prime }=  \notag \\
&=&1+\alpha ^{2}x^{2}+\beta ^{2}p^{2}+2\alpha \beta (xp)  \notag \\
&&+\left[ 2\alpha ^{2}\left( 1+\alpha ^{2}x^{2}+\beta ^{2}p^{2}+2\alpha
\beta \left( xp\right) \right) x_{j}+2\alpha \beta \left( 1+\alpha
^{2}x^{2}+\beta ^{2}p^{2}+2\alpha \beta xp\right) p_{j}\right] \frac{%
\partial H}{\partial p_{j}}\delta t+  \notag \\
&&+\left[ -2\beta ^{2}\left( 1+\alpha ^{2}x^{2}+\beta ^{2}p^{2}+2\alpha
\beta \left( xp\right) \right) p_{j}-2\alpha \beta \left( 1+\alpha
^{2}x^{2}+\beta ^{2}p^{2}+2\alpha \beta x\cdot p\right) x_{j}\right] \frac{%
\partial H}{\partial x_{j}}\delta t+O\left( \delta t^{2}\right)   \notag \\
&=&\left( 1+\alpha ^{2}x^{2}+\beta ^{2}p^{2}+2\alpha \beta (xp)\right)
[1+\left( 2\alpha ^{2}x_{j}+2\alpha \beta p_{j}\right) \frac{\partial H}{%
\partial p_{j}}\delta t-\left( 2\beta ^{2}p_{j}+2\alpha \beta x_{j}\right) 
\frac{\partial H}{\partial x_{j}}\delta t]+O\left( \delta t^{2}\right) . 
\notag \\
&&
\end{eqnarray}%
We can see that the weighted volume element will stay invariant
since the weight factor we introduced produces the same terms as the
Jacobian in \eqref{jacobian} under the infinitesimal time evolution (up to
the first order in $\delta t$): 
\begin{eqnarray}
&&\frac{d^{D}x^{\prime }d^{D}p^{\prime }}{1+\alpha ^{2}x^{\prime 2}+\beta
^{2}p^{\prime 2}+2\alpha \beta x^{\prime }p^{\prime }}= \nonumber\\
&=&\frac{d^{D}xd^{D}p\left( 1+2\left[ \left( \alpha ^{2}x_{j}+\alpha \beta
p_{j}\right) \frac{\partial H}{\partial p_{j}}-\left( \beta ^{2}p_{j}+\alpha
\beta x_{j}\right) \frac{\partial H}{\partial x_{j}}\right] \delta t+O\left(
\delta t^{2}\right) \right) }{\left( 1+\alpha ^{2}x^{2}+\beta
^{2}p^{2}+2\alpha \beta x\cdot p\right) \left( 1+2\left[ \left( \alpha
^{2}x_{j}+\alpha \beta p_{j}\right) \frac{\partial H}{\partial p_{j}}-\left(
\beta ^{2}p_{j}+\alpha \beta x_{j}\right) \frac{\partial H}{\partial x_{j}}%
\right] \delta t+O\left( \delta ^{2}\right) \right) } \nonumber\\
&=&\frac{d^{D}xd^{D}p}{\left( 1+\alpha ^{2}x^{2}+\beta ^{2}p^{2}+2\alpha
\beta xp\right) }.
\end{eqnarray}%
We point out this result holds to any order of parameters $\alpha $ and $%
\beta $ (as we have not done any expansions in the noncommutative
parameters). 

\vspace{0.5cm} 

We note that from the SdS model considered above we can
obtain two well known special cases. Namely: 

\begin{itemize}
\item Snyder model \cite{snyder1947quantized}\footnote{
Snyder model was the first proposed noncommutative space-time model preserving Lorentz symmetry.} is obtained when we take $\alpha \rightarrow 0$ in SdS algebra 
\eqref{xxTSR}-\eqref{xpTSR}, i.e. we obtain the model with noncommutative coordinates and
commutative momenta (curved momentum space): 
\begin{equation}
\lbrack \hat{x}_{\mu },\hat{x}_{\nu }]=i\hbar \beta ^{2}\hat{M}_{\mu \nu
},\qquad \lbrack \hat{p}_{\mu },\hat{p}_{\nu }]=0,\qquad \lbrack \hat{x}%
_{\mu },\hat{p}_{\nu }]=i\hbar (\eta _{\mu \nu }+\beta ^{2}\hat{p}_{\mu }%
\hat{p}_{\nu }),  \label{Snyder}
\end{equation}%
supplemented by the Lorentz covariance conditions \eqref{Mx} - \eqref{MM}.
Such modification of quantum mechanical phase space relations leads, in the
non-relativistic case, to the quadratic GUP (QGUP) \footnote{for the simple case in which $\left\langle \hat{p}_{i}\right\rangle =0.$}: 
\begin{equation}
\Delta \hat{x}_{i}\Delta \hat{p}_{i}\geq \frac{\hbar }{2}(1+\beta ^{2}(\Delta \hat{p})^{2}).
\label{qgup}
\end{equation}%
We note that the algebraic set of commutation relations \eqref{Snyder} is only one of the
many possible realizations of the Snyder model and more general realizations
may be considered, see e.g. \cite{Pachol:2023tqa,Pachol:2023bkv}. 
The non-relativistic classical Poisson algebra: 
\begin{equation}
\{x_{i},x_{j}\}=\beta ^{2}(x_{i}p_{j}-x_{j}p_{i}),\qquad
\{p_{i},p_{j}\}=0,\qquad \{x_{i},p_{j}\}=\delta _{ij}+\beta ^{2}p_{i}p_{j}\label{PoisSnyder}
\end{equation}%
will result in the following weighted phase space volume element: 
\begin{equation}
\frac{d^{D}xd^{D}p}{1+\beta ^{2}p^{2}}  \label{dxdpSnyder}
\end{equation}
(in any dimension $D$), obtained as the limit $\alpha \rightarrow 0$ in %
\eqref{dxdp_inv}. 
\item (Anti-)de Sitter model (dual Snyder model) is obtained  when we take $\beta
\rightarrow 0$ in SdS algebra \eqref{xxTSR}-\eqref{xpTSR}, i.e. we obtain the model with commutative
coordinates but noncommutative momenta (curved space-time \footnote{We recall, that the parameter $\alpha $ still will play the role of the inverse of the de Sitter radius $R$ i.e. space-time curvature, which is linked to the cosmological constant.}): 
\begin{equation}
\lbrack \hat{x}_{\mu },\hat{x}_{\nu }]=0,\qquad \lbrack \hat{p}_{\mu },\hat{p}_{\nu }]=i\hbar \alpha ^{2}\hat{M}_{\mu \nu },\qquad \lbrack \hat{x}_{\mu },\hat{p}_{\nu }]=i\hbar \left( \eta _{\mu \nu }+\alpha ^{2}\hat{x}_{\mu }\hat{x}_{\nu }\right) .  \label{xp_cdS}
\end{equation}
with the Lorentz covariance given by \eqref{Mx} - \eqref{MM}. The non-relativistic case results in the quadratic EUP (QEUP) \footnote{for the simple case in which $\left\langle\hat{x}_{i}\right\rangle =0.$}: 
\begin{equation}
\Delta \hat{x}_{i}\Delta \hat{p}_{i}\geq \frac{\hbar }{2}(1+\alpha ^{2}(\Delta \hat{x})^{2}).
\label{qeup}
\end{equation}%
Recently, in \cite{Ghosh:2024eza} it has been shown that such EUP can appear in canonical quantum mechanics, assuming the periodic nature in coordinate or momentum space.

The non-relativistic classical Poisson algebra is: 
\begin{equation}
\{x_{i},x_{j}\}=0,\qquad \{p_{i},p_{j}\}=\alpha
^{2}(x_{i}p_{j}-x_{j}p_{i}),\qquad \{x_{i},p_{j}\}=\delta _{ij}+\alpha
^{2}x_{i}x_{j}.
\end{equation}
And the weighted phase space volume element (in any dimension $D$), obtained as the
limit $\beta \rightarrow 0$ in \eqref{dxdp_inv}, is: 
\begin{equation}
\frac{d^{D}xd^{D}p}{1+\alpha ^{2}x^{2}}.
\end{equation}
\end{itemize}
It is worth to point out 
that there exists a way to transform SdS algebra \eqref{xxTSR}-\eqref{xpTSR} generated by ($\hat{x},\hat{p},\hat{M}$) into the Snyder algebra \eqref{Snyder} generated by ($\hat{x}^S,\hat{p}^S,\hat{M}$) by the following linear maps: 
\begin{equation}
\hat{x}_{i}=\hat{x}^S_{i}+\frac{\beta}{\alpha }\lambda \hat{p}^S_{i},\quad \hat{p}_{i}=\left( 1-\lambda \right) \hat{p}^S_{i}-\frac{\alpha }{\beta }\hat{x}^S_{i}
\end{equation}
where $\lambda $ is a free parameter and we have temporarily denoted the Snyder algebra \eqref{Snyder} generators by the upper index $S$. This relation of the SdS algebra with the Snyder algebra was first presented in \cite{Mignemi:2011wh}.
Through such noncanonical change of basis one can use the already known machinery of realizations in Snyder spaces and
consider various applications of SdS algebra, for example to find harmonic oscillator solutions \cite{Mignemi:2011gr} or in applications to quantum field theory \cite{Franchino-Vinas:2024ltq}.

One can also consider various generalizations of the SdS model. For example,
in \cite{Bilac:2024xxq} the following generalization of the last relation in
SdS algebra 
\eqref{xpTSR} was proposed: 
\begin{equation}
\lbrack \hat{x}_{\mu },\hat{p}_{\nu }]=i\hbar \left( \eta _{\mu \nu }\varphi
_{1}+\left( \alpha ^{2}\hat{x}_{\mu }\hat{x}_{\nu }+\beta ^{2}\hat{p}_{\mu }%
\hat{p}_{\nu }+\alpha \beta \hat{x}_{\mu }\hat{p}_{\nu }+\alpha \beta \hat{p}%
_{\mu }\hat{x}_{\nu }\right) \varphi _{2}-\alpha \beta \hat{M}_{\mu \nu
}\right)   \label{xpTSR-gen}
\end{equation}%
where the functions $\varphi _{1}$ and $\varphi _{2}$ need to satisfy
specific conditions (due to Jacobi identities). When $\varphi _{1}=\varphi
_{2}=1$ we get back the original SdS model \eqref{xpTSR}. Various choices of 
$\varphi _{1},\ \varphi _{2}$ are discussed in \cite{Bilac:2024xxq} and for
example for one specific choice of $\varphi _{1}$ and $\varphi _{2}$ we can
obtain the following relation: 
\begin{equation}
\lbrack \hat{x}_{\mu },\hat{p}_{\nu }]=i\hbar \eta _{\mu \nu }\sqrt{1-\left(
\alpha ^{2}\hat{x}^{2}+\beta ^{2}\hat{p}^{2}+\alpha \beta \hat{x}\hat{p}%
+\alpha \beta \hat{p}\hat{x}\right) }-\alpha \beta \hat{M}_{\mu \nu }.
\label{xpTSR-realization}
\end{equation}%
Such generalizations would be interesting to investigate further in the
context of GEUPs and their influence on the density of states. 
\vspace{0.5cm} 

Before completing this section, since the effects of modifications of UPs on
the density of states have been investigated previously few comments
are in order. In \cite{Chang:2001bm,Chang:2001kn}, in the case of GUP %
\eqref{qgup}\footnote{In \cite{Chang:2001bm,Chang:2001kn} more general case, than \eqref{qgup}, is considered with two parameters $\beta$, $\beta ^{\prime }$. In the
comparison we take $\beta ^{\prime }=0$ since this is the most similar
option to the Snyder model discussed here. Nevertheless, the case with $%
\beta ^{\prime }\neq 0$ can also be associated with the Snyder model, but
requires different (more general) realization 
see
e.g. \cite{Pachol:2023tqa,Pachol:2023bkv}.} the invariant weighted
phase space volume element is obtained as: $\frac{d^{D}xd^{D}p}{\left(
1+\beta^2 p^{2}\right) ^{D}}$, where the power $D$ (dimension) is necessary
since the Jacobian obtained is: $d^{D}x^{\prime }d^{D}p^{\prime
}=d^{D}xd^{D}p\left( 1-2\beta^2 Dp_{k}\frac{\partial H}{\partial x_{k}}\delta
t+O\left( \delta t^{2}\right) \right) $. This is different than the case
considered here, where in the limit $\alpha\to 0$ we obtain $\frac{d^{D}xd^{D}p}{1+\beta ^{2}p^{2}}$  \eqref{dxdpSnyder}. The difference arises from the
fact that commutation relations used in \cite{Chang:2001bm} for $[x_i,p_j]$
have the term proportional to $p^2$ while here we have the terms with $%
p_ip_j $ instead, cf. \eqref{PoisSnyder}. In the case of the Snyder model
with \eqref{PoisSnyder} the terms with $D$ in the Jacobian \eqref{jacobian} cancel out, hence the power $D$
is not appearing in the weight factor of the volume form \eqref{dxdpSnyder}. 

Other important point is that often in the context of modified
UP \eqref{UP} also the inner product on the Hilbert space\footnote{
The Heisenberg algebra \eqref{genCRhat}, \eqref{genCRxp} is represented on
the space of states in which one usually chooses a basis of position or
momentum eigenvectors. In the case of GUP one usually chooses
the momentum space, while in the case of GEUP Bergmann-Fock construction can be used \cite{Kempf:1993bq}.}
becomes modified with appropriately chosen measure so
that the observables (satisfying the modified commutation relations) stay
symmetric on the dense domain of functions decaying faster than any power 
\cite{Kempf:1994su}. For the inner product modifications in the case of GUP see, e.g. \cite{Chang:2001bm,Chang:2001kn} and for the relativistic case, see e.g. \cite{Quesne:2006fs}. The inner product modifications in the
case of EUP, i.e. in AdS and dS spaces were investigated e.g. in \cite%
{Hamil:2019pum}. Subsequently the modified inner product can be used to
study effects on solutions of Schr\"odinger equation, see e.g. \cite%
{Chang:2001bm,Chang:2001kn,Hamil:2019pum}. Such modifications in the measure
in the inner product have not been under investigation of the present paper.

\section{Yang model}

The Yang model introduced in \cite{Yang:1947ud} is a Lorentz
invariant model incorporating noncommutative space-time coordinates as well
as noncommutative momenta, depending on the pair of dimensionful parameters $\alpha $ and $\beta $ related with the curvatures of quantum space-time and
momentum spaces (in similarity to SdS model). The defining relations are as
follows: 
\begin{equation}
\lbrack \hat{x}_{\mu },\hat{x}_{\nu }]=i\hbar \beta ^{2}\hat{M}_{\mu \nu
},\qquad \lbrack \hat{p}_{\mu },\hat{p}_{\nu }]=i\hbar \alpha ^{2}\hat{M}%
_{\mu \nu }  \label{xxYang}
\end{equation}%
\begin{equation}
\lbrack \hat{M}_{\mu \nu },\hat{x}_{\rho }]=i\hbar (\eta _{\mu \rho }\hat{x}
_{\nu }-\eta _{\nu \rho }\hat{x}_{\mu }),\label{xMYang}
\end{equation}
\begin{equation}
\lbrack \hat{M}_{\mu \nu },\hat{p}_{\rho }]=i\hbar (\eta _{\mu \rho }\hat{p}
_{\nu }-\eta _{\nu \rho }\hat{p}_{\mu }),\label{pMYang}
\end{equation}%
\begin{equation}
\lbrack \hat{M}_{\mu \nu },\hat{M}_{\rho \tau }]=i\hbar (\eta _{\mu \rho }%
\hat{M}_{\nu \tau }-\eta _{\mu \tau }\hat{M}_{\nu \rho }+\eta _{\nu \tau }%
\hat{M}_{\mu \rho }-\eta _{\nu \rho }\hat{M}_{\mu \tau }).\label{MMYang}
\end{equation}%
However, the quantum phase space relation is described by an additional generator $%
\hat{r}$ (central charge): 
\begin{equation}
\lbrack \hat{x}_{\mu },\hat{p}_{\nu }]=i\hbar \eta _{\mu \nu }\hat{r},
\label{xpYang}
\end{equation}%
hence to obtain the full Yang algebra we need the additional relations: 
\begin{equation}
\lbrack \hat{r},\hat{x}_{\mu }]=i\hbar \beta ^{2}\hat{p}_{\mu },\qquad
\lbrack \hat{r},\hat{p}_{\mu }]=-i\hbar \alpha ^{2}\hat{x}_{\mu },\qquad
\lbrack \hat{M}_{\mu \nu },\hat{r}]=0.  \label{rM}
\end{equation}%
The uncertainty principle corresponding to the Yang model, in the
non-relativistic case, can be written in general as: 
\begin{equation}
\Delta \hat{x}_{i}\Delta \hat{p}_{j}\geq \frac{\hbar \delta _{ij}}{2}\left\vert \langle 
\hat{r}\rangle \right\vert .  \label{rup}
\end{equation}%
where the generator $\hat{r}$ can be realized in terms of the
phase space variables $\hat{r}=\hat{r}\left(\hat{x},\hat{p}\right) $.  It
is also worth to mention that the Yang model is covariant (self-dual) under
the Born reciprocity: 
\begin{equation}
B:\quad \hat{x}_{\mu }\rightarrow \hat{p}_{\mu },\quad \hat{p}_{\mu }\rightarrow -\hat{x}%
_{\mu },\quad \hat{M}_{\mu \nu }\leftrightarrow \hat{M}_{\mu \nu },\quad 
\hat{r}\leftrightarrow \hat{r},\quad \alpha \leftrightarrow \beta .
\label{born}
\end{equation}%
The similarity between the Yang model and the SdS model considered in the previous
section is not coincidental and it has been shown \cite%
{Chryssomalakos:2004wc} that the SdS algebra \eqref{xxTSR}-\eqref{xpTSR} can be viewed
as a nonlinear realization of the Yang model \eqref{xxYang}-\eqref{rM}.

The
classical limit of the Yang model (cf. \cite%
{Meljanac:2023dpr}, see also \cite{Bilac:2024wex}) is: 
\begin{equation}
\{{x}_{i},{x}_{j}\}=\beta ^{2}(x_{i}p_{j}-x_{j}p_{i}),\qquad \{{p}_{i},{p}%
_{j}\}=\alpha ^{2}(x_{i}p_{j}-x_{j}p_{i})  \label{xxnr}
\end{equation}%
and one of the possible realizations for the $\hat{r}$ generator on the
canonical phase space \cite{Meljanac:2023dpr}, for example, gives: 
\begin{equation}
\{x_{i},{p}_{j}\}=\delta _{ij}\sqrt{1-\alpha ^{2}{x}^{2}-\beta
^{2}p^{2}-\alpha ^{2}\beta ^{2}(x^{2}p^{2}-(xp)^{2})}.  \label{xpnr}
\end{equation}%
The remaining relations are obtained in a straightforward way. 
In D=1 this would simplify to: 
\begin{equation}
\{x,{p}\}=\sqrt{1-\alpha ^{2}{x}^{2}-\beta ^{2}p^{2}}. \label{xpYP-1}
\end{equation}%
with the corresponding GEUP \footnote{In higher dimensions additional mixed terms would be present in the GEUP relation.
}: 
\begin{equation}
\Delta x\Delta p\geq \frac{\hbar }{2}\sqrt{1-\alpha ^{2}(\Delta x)^{2}-\beta
^{2}(\Delta p)^{2}}. \label{regup}
\end{equation}
We see that in such 1-dimensional case higher order terms would appear, which is not uncommon for GUPs see e.g. \cite{Nouicer:2007jg,Chung:2019raj,Fadel:2021hnx} but this would be the first such example considering higher order GEUPs, up to our knowledge.

Expanding the RHS of \eqref{xpYP-1} in
deformation parameters, up to $\alpha ^{2}$ and $\beta ^{2}$, we can
apply the results of the previous section \eqref{dxdp_inv}
and obtain the invariant weighted phase space volume element as: 
\begin{equation}
\frac{dxdp}{1-\frac{1}{2}\left( \alpha ^{2}x^{2}+\beta ^{2}p^{2}\right) }.
\label{dxdp_invYP1}
\end{equation}
From the Yang model \eqref{xxYang}-\eqref{rM}, in the realization \eqref{xpnr} for $\hat{r}$ generator we can
obtain two special cases. Namely: 
\begin{itemize}
\item In the limit when $\alpha \rightarrow 0$ we obtain the so-called "square-root modified" or "Maggiore algebra" (see e.g. \cite{Battisti:2008xy}):
\begin{equation}
\lbrack \hat{x}_{\mu },\hat{x}_{\nu }]=i\hbar \beta ^{2}\hat{M}_{\mu \nu
},\qquad \lbrack \hat{p}_{\mu },\hat{p}_{\nu }]=0,\qquad \lbrack \hat{x}%
_{\mu },\hat{p}_{\nu }]=i\hbar \eta _{\mu \nu }\sqrt{1-\beta ^{2}
 \hat{p} ^{2}},  \label{Yang_root}
\end{equation}%
supplemented by the Lorentz covariance conditions \eqref{xMYang} - \eqref{MMYang}.
Such modification of quantum mechanical phase space relations leads, in the
non-relativistic case, to the higher order type of GUP \footnote{We call this type of GUP "higher order" due to the expansion of the square root function in powers of $p$ thanks to which, through series of inequalities, one obtains the formula \eqref{sqrtUP}, see e.g.\cite{Fadel:2021hnx}.} of the form:
\begin{equation}
\Delta \hat{x}_i\Delta \hat{p}_i\geq \frac{\hbar }{2}\sqrt{1-\beta ^{2}\left(
\Delta  \hat{p}\right) ^{2}}.\label{sqrtUP}
\end{equation}
This version of GUP no longer produces a minimum observable length \cite{Fadel:2021hnx,Segreto:2022clx}. But it would still result in the weighted phase space volume element:
\begin{equation}
\frac{dxdp}{1-\frac{1}{2}\beta ^{2}p^{2} }
\label{dxdp_inv_sqrt}
\end{equation}
affecting the density of states.
\item By taking $\beta \rightarrow 0$ in \eqref{xxYang}-\eqref{rM}, we obtain
\begin{equation}
\lbrack \hat{x}_{\mu },\hat{x}_{\nu }]=0,\qquad \lbrack \hat{p}_{\mu },\hat{p}_{\nu }]=i\hbar \alpha ^{2}\hat{M}_{\mu \nu
},\qquad \lbrack \hat{x}%
_{\mu },\hat{p}_{\nu }]=i\hbar \eta _{\mu \nu }\sqrt{1-\alpha ^{2}
 \hat{x} ^{2}},  \label{Yang_root}
\end{equation}
leading to the higher order type of EUP \footnote{By similar expansion of the square-root function in powers of $x$.}:
\begin{equation}
\Delta \hat{x}_i\Delta \hat{p}_i\geq \frac{\hbar }{2}\sqrt{1-\alpha ^{2}\left(
\Delta  \hat{x}\right) ^{2}}
\end{equation}
with the weighted phase space volume element as:
\begin{equation}
\frac{dxdp}{1-\frac{1}{2}\alpha ^{2}x^{2} }
\label{dxdp_inv_sqrt2}
\end{equation}
affecting the density of states.
\end{itemize}

We postpone the investigation of the full D-dimensional case of Yang model \eqref{xxYang}-\eqref{rM} in the context of density of states to the future work. It is also worth to mention that in \cite{Meljanac:2022qhp}
generalizations of the Snyder algebra to a curved space-time background with
de Sitter symmetry were considered where the SdS model and Yang model were
obtained as special cases. The realizations of these algebras were
considered in terms of canonical phase space coordinates, up to the fourth order
in the deformation parameters. Therefore the results of the present paper
could be generalized to these type of models and realizations as well.

\section{Lie-algebraic case with commuting momenta: Fuzzy sphere}
Many noncommutative (quantum) space-times proposed in the quantum gravity motivated literature have the Lie-algebraic form for the noncommutativity of coordinates and include commuting momenta.
Hence, for the sake of completeness we discuss here one example of such model defined by the following commutation relations:
\begin{equation}
\left[ \hat{x}_{i},\hat{x}_{j}\right] =i\hbar \epsilon _{ijk}\hat{x}_{k},\qquad \left[
\hat{x}_{i},\hat{p}_{j}\right] =i\hbar \epsilon _{ijk}\hat{p}_{k},\qquad \left[ \hat{p}_{i},\hat{p}_{j}\right] =0
\end{equation}
where $\epsilon _{ijk}$ is totally skew-symmetric tensor, $\epsilon _{123}=1.$ The subalgebra $[\hat{x}_{i},\hat{x}_{j}]=i\hbar \epsilon _{ijk}\hat{x}_{k}$ supplemented by the relation $\hat{x}^i\hat{x}_i=r$ corresponds to the fuzzy sphere, with $r$ the constant radius of the sphere.

Since we are interested in the effects on the density of states in this Lie algebraic case we follow the steps outlined in Sec.2. In the classical limit, we obtain
: 
\begin{equation}
\{x_{i},x_{j}\}=\epsilon _{ijk}x_{k}\equiv a_{ij},\qquad \{x_{i},p_{j}\}=\epsilon
_{ijk}p_{k}\equiv c_{ij},\qquad \{p_{i},p_{j}\}=0\equiv b_{ij},
\end{equation}
where we identified RHSs with the notation used in Sec. 2. Directly plugging these relations into \eqref{dxdp} we obtain:
\begin{eqnarray}
d^{D}x^{\prime }d^{D}p^{\prime } &=&d^{D}xd^{D}p\left( 1+\left[ \left( \frac{%
\partial }{\partial x_{i}}\left( \epsilon _{ijk}x_{k}\right) -\frac{\partial 
}{\partial p_{i}}\left( \epsilon _{jik}p_{k}\right) \right) \frac{\partial H%
}{\partial x_{j}}+\left( \frac{\partial }{\partial x_{i}}\left( \epsilon
_{ijk}p_{k}\right) \right) \frac{\partial H}{\partial p_{j}}\right] \delta
t+O\left( \delta t^{2}\right) \right) \nonumber \\
&=&d^{D}xd^{D}p\left( 1+\left[ \left( \epsilon _{ijk}\delta _{ik}-\epsilon
_{jik}\delta _{ik}\right) \frac{\partial H}{\partial x_{j}}+0\right] \delta
t+O\left( \delta t^{2}\right) \right) =d^{D}xd^{D}p.
\end{eqnarray}
Hence we see that the infinitesimal phase space volume element stays invariant under the time evolution
without the need to introduce any additional factor
and there will be no change in the density of states for the case of fuzzy sphere.

\section{Final remarks}

 In this paper we have focused on models exhibiting
noncommutativity in both space-time coordinates and in momenta (i.e. models
with curved space-time and curved momentum space), such that in the quantum phase
space relations both the Planck length $l_P$ and the cosmological constant $\Lambda$ appear as fundamental parameters on equal footing. The modified
Heisenberg commutation relations lead to GEUPs where the symmetry between
position and momentum is preserved (which is not the case in the usual GUPs
or EUPs). Such symmetric GEUPs may be seen as an indication of quantum
gravitational corrections to the classical space-time and standard quantum mechanics, at both very small and very
large scales. 
We point out that, on the contrary to many  works which have proposed multiple variants of the GUPs, EUPs and GEUPs by arbitrarily choosing specific forms
of the commutation relations between space-time coordinates and momenta $[x,p]$  with supposedly desirable properties, we have focued here on studying the consequences of known models which arise in the noncommutative geometry approach to quantum gravity.

In general since the canonical commutation relations are
modified, one expects that thermodynamics and statistical mechanics will be
affected by the introduced modifications, possibly leading to some new
effects. As a consequence of the GEUP arising from the cases of Snyder-de
Sitter and Yang models we have shown that the analogue of the Liouville theorem in
statistical physics requires considering the weighted phase space volume and
introduces modification in the density of states, with the weight factor
depending on both coordinates and momenta. Since such modification
of the density states is required this will influence the statistical and thermodynamical
properties of physical systems. Various applications can now be studied and
the effects of both noncommutativity in coordinates and momenta (or the
presence of the Planck length and the cosmological constant in modified UPs) on
atomic physics, condensed matter physics, preheating phase of the universe
and black holes etc. can now be investigated. 

\subsection*{Acknowledgements}

AP thanks A. Wojnar and S. Zonetti for interesting discussions
and acknowledges the support of the Polish NCN grant 2022/45/B/ST2/01067 and
COST Action CA21109 - CaLISTA .

\end{document}